\documentclass[aps,prb,superscriptaddress,twocolumn,showpacs,nofootinbib,amsmath]{revtex4}
\usepackage{graphicx}

\newcommand{\be}{\begin{equation}}
\newcommand{\ee}{\end{equation}}
\newcommand{\bse}{\begin{subequations}}
\newcommand{\ese}{\end{subequations}}
\newcommand{\bea}{\begin{eqnarray}}
\newcommand{\eea}{\end{eqnarray}}
\newcommand{\fra}[2]{\hbox{${#1\over #2}$}}
\newcommand{\comment}[1]{}

\begin{document}

\title{
Even-odd effects in finite Heisenberg spin chains}

\author{Paolo Politi}
\email{paolo.politi@isc.cnr.it}
\affiliation{Istituto dei Sistemi Complessi,
Consiglio Nazionale delle Ricerche, Via Madonna del Piano 10,
50019 Sesto Fiorentino, Italy}

\author{Maria Gloria Pini}
\email{mariagloria.pini@isc.cnr.it}
\affiliation{Istituto dei Sistemi Complessi,
Consiglio Nazionale delle Ricerche, Via Madonna del Piano 10,
50019 Sesto Fiorentino, Italy}

\date{\today}

\begin{abstract}
Magnetic superlattices and nanowires may be described
as Heisenberg spin chains of finite length $N$, where $N$ is the number of magnetic units 
(films or atoms, respectively). We study antiferromagnetically coupled spins which are also
coupled to an external field $H$ (superlattices) or to a ferromagnetic substrate (nanowires).
The model is analyzed through a two-dimensional map which allows fast and reliable 
numerical calculations. Both open and closed chains have different properties
for even and odd $N$ (parity effect).
Open chains with odd $N$ are known [S.~Lounis {\it et al.}, Phys. Rev. Lett.
{\bf 101}, 107204 (2008)] to have a ferrimagnetic state for small $N$ 
and a noncollinear state for large $N$. 
In the present paper, the transition length $N_c$ is found analytically. 
Finally, we show that closed chains arrange themselves in the uniform bulk spin-flop state for even $N$
and in nonuniform states for odd $N$.
\end{abstract}


\maketitle

Antiferromagnetic (AF) chains of Heisenberg spins, when subjected
to an external magnetic field (and possibly to a uniaxial
anisotropy), are known to arrange themselves in a spin-flop (SF) state
where neighboring spins are almost antiparallel and orthogonal to
the field~\cite{Neel}. However, such a result is valid, strictly speaking,
only in the thermodynamic limit. For a finite chain, boundary
conditions and finite size effects are expected to induce
modifications on the bulk spin-flop configuration, determining a
non uniform canting along the chain. 

A one-dimensional (1D) classical planar model of a Heisenberg uniaxial
antiferromagnet,
\begin{eqnarray}\label{uniaxial}
{\cal H}_{AF}&=&\sum_{i=1}^{N-1}H_E \cos(\theta_{i}-\theta_{i+1})\cr
&-&\sum_{i=1}^N (H_A \cos^2\theta_i+2H \cos\theta_i),
\end{eqnarray}
was introduced forty years ago~\cite{Mills} to study a semi-infinite
AF chain ($N\to\infty$).
In Eq.~(\ref{uniaxial}), $H_E$ denotes
the exchange field, $H_A$ the anisotropy field, and $\theta_i$ is
the angle that the magnetization of the $i$-th ferromagnetic layer
forms with the direction of the external field, $H$ (spins are assumed
to be planar). The bulk SF phase appears for $H>\sqrt{2H_EH_A+H_A^2}$
($H>0$ for zero anisotropy $H_A$).

At that time, the reference experimental systems were bulk systems like
MnF$_2$ or MnO, with magnetic ions on special crystallographic
planes interacting ferromagnetically (FM) and ions between planes
interacting AF. With the spreading of epitaxially grown systems,
the model was used a lot
\cite{WangMillsPRL,WangMillsPRB,PRL,JAP,IJMPB,Bogdanov}
to study superlattices made of $N$ ferromagnetic layers which are
antiferromagnetically coupled. 
Some important theoretical results were found: (i) For semi-infinite systems, 
the surface SF state (a phase predicted to anticipate the bulk SF state when
increasing the field) does not exist~\cite{PRL}; 
(ii) For finite systems, there are important differences
between structures with even and odd $N$~%
\cite{WangMillsPRL,WangMillsPRB,JAP,IJMPB}.

Recently, the model of a finite 1D quantum Heisenberg
antiferromagnet
\begin{equation}
\label{quantum} {\cal H}_q=\vert J_1 \vert \sum_{i=1}^{N-1} {\bf
S}_i \cdot {\bf S}_{i+1}
\end{equation}
has gained new interest since it has been used to describe an AF
nanowire deposited on a thin {\it insulating} layer
\cite{Science}. Paradigmatic examples of such a system are linear
chains of 1 to 10 Mn atoms epitaxied on a CuN substrate
\cite{Science}. From the analysis of spin excitations of coupled
atomic spins in the dimer and in the trimer ($N=2,3$), the Mn-Mn
exchange interaction was found \cite{Science} to be
antiferromagnetic ($\vert J_1 \vert =6.2$ meV) and the spin value to be
$S=5/2$, identical to the spin of a free Mn atom. Using these
parameters in Eq.~(\ref{quantum}), the magnetic behavior of longer
wires could successfully be fitted \cite{Science}.

When such AF Mn nanowires are deposited on a {\it ferromagnetic}
layer, like Ni(001), an interesting frustration phenomenon occurs,
since the exchange coupling between an adsorbed Mn spin and the
magnetic moment of an underlying Ni atom of the substrate is
ferromagnetic \cite{LounisPRB,LounisPRL}, $J_2>0$, and thus
competes \cite{notaLounis} with the Mn-Mn antiferromagnetic
exchange, $J_1<0$. Therefore, in a classical spin approximation,
one is led to consider the model~\cite{LounisPRL}
\begin{equation}
\label{competing} {\cal H}_N=\vert J_1 \vert \sum_{i=1}^{N-1}
\cos(\theta_i-\theta_{i+1})-J_2 \sum_{i=1}^N \cos\theta_i ,
\end{equation}
\noindent where $\theta_i$ denotes
the angle that the $i$-th spin of the AF nanowire forms with
respect to the magnetization of the ferromagnetic substrate. Since
the coupling $J_2$ is localized on the $i$-th site, it is apparent
that it plays the same role as the magnetic field $2H$ in
Eq.~(\ref{uniaxial}), while $\vert J_1 \vert$ has to be identified
with $H_E$, and $H_A$ (the uniaxial anisotropy) is zero. Therefore, when $N$ is finite
and open boundary conditions are assumed,  the existence of
different ground states for odd $versus$ even $N$ is a well-known
result \cite{WangMillsPRL,IJMPB}. Different ground states also
reflect on different behaviors for the spin wave excitations
\cite{WangMillsPRB,JAP}.

In a recent Letter~\cite{LounisPRL} S. Lounis {\it et al.}, using
both {\it ab initio} results and solutions to the classical
Heisenberg model (\ref{competing}), confirmed that the ground state
of finite AF nanowires deposited on ferromagnets depends on the
parity of the number $N$ of atoms. They also found that, while even
chains always have a noncollinear (NC) ground state, for odd $N$ a transition
from a collinear ferrimagnetic (FI) to a NC
configuration occurs when the chain length $N$ exceeds a critical
value $N_c$. For example, using an iterative numerical scheme 
in order to minimize Eq.~(\ref{competing}), the transition length was
estimated \cite{LounisPRL} to be 9 atoms for Mn chains on Ni(001).

Here we show that the classical Heisenberg model
(\ref{competing}) can be investigated with great numerical
and analytical profit in terms of a
two-dimensional (2D) map method \cite{PRL,JAP,IJMPB,Aubry,Bak,Belorov,Pandit}. 
Such an
approach allows a fast and exact determination of the ground state
configuration of finite chains and to find an {\it
analytical} expression for the transition length for odd open chains.

\begin{figure}
\includegraphics[width=8cm,clip=yes]{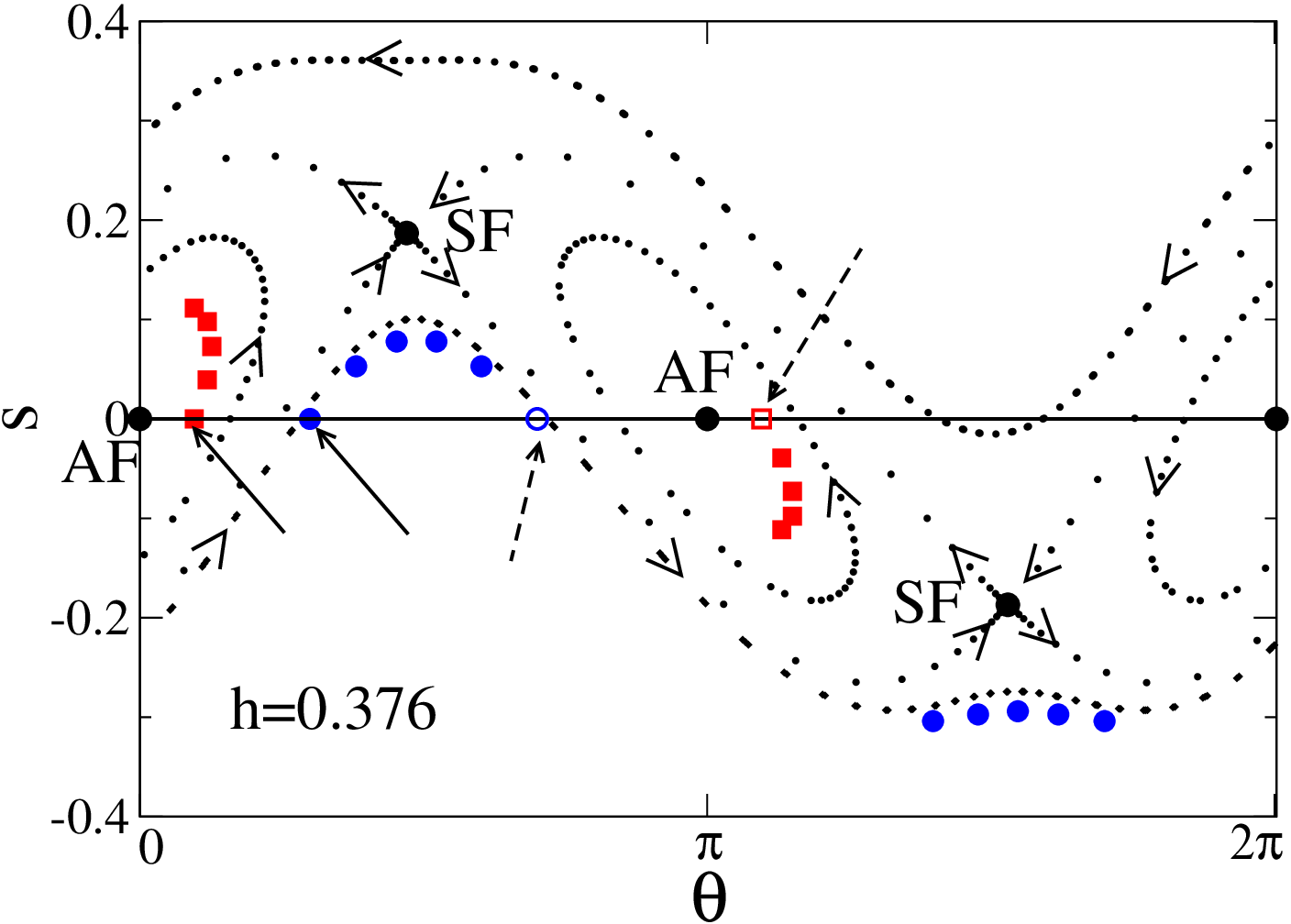}
\caption{ (color online) Phase portrait for mapping (\ref{mappa}).
Full red squares and blue circles correspond to
equilibrium configurations for open chains with $N=9$ and $N=10$,
respectively. Full arrows point to the values $\theta_1$ for the
first atom of the chains. Dashed arrows point to the values
$\theta_{N+1}$ (denoted by open symbols) for the fictitious
$(N+1)$-th atom. Arrow tips allow one to follow the evolution of
the map, whose hyperbolic and elliptic fixed points of order two,
respectively denoted by SF and AF, correspond to uniform states. }
\label{fig_map}
\end{figure}

By the map method, we
also study model (\ref{competing}) in the case of periodic
boundary conditions. For such ``closed" chains, we find a new
even-odd effect: even chains have spins arranged in the spin-flop
state, like infinite chains, while odd chains arrange themselves 
in noncollinear states.

In order to find the equilibrium configurations of the classical
Heisenberg model (\ref{competing}) by the 2D map method,
we introduce~\cite{Belorov} the variable $s_n=\sin(\theta_n
- \theta_{n-1})$. Denoting by $h=J_2/|J_1|$ the ratio between
competing exchange interactions, minimization of (\ref{competing})
gives~\cite{note_sin}
\begin{equation}
\label{mappa} s_{n+1} = s_n - h\sin\theta_n, \hbox{~~~}
\theta_{n+1} = \theta_n + \sin^{-1}(s_{n+1}) .
\end{equation}
These equations define an iterative 2D map~\cite{Ott},
i.e. point $(\theta,s)$ in the phase space is mapped to
a point $(\theta',s')$. The fixed points of
order two ($s_{n+2}=s_n$ and $\theta_{n+2}=\theta_n$) correspond
to the collinear AF configuration
($(0,0)\leftrightarrow (\pi,0)$) and to the bulk SF state
($(\bar\theta,\sin 2\bar\theta)\leftrightarrow (-\bar\theta,-\sin
2\bar\theta)$, with $\cos\bar\theta = h/4$). In Fig.~\ref{fig_map}
we plot the fixed points and the evolution of the map for different initial conditions
and $h=0.376$ (it is the special value considered in
Ref.~\onlinecite{LounisPRL} as representative of AF Mn nanowires
on Ni(001)).

\begin{figure}
\includegraphics[width=8cm,clip=yes]{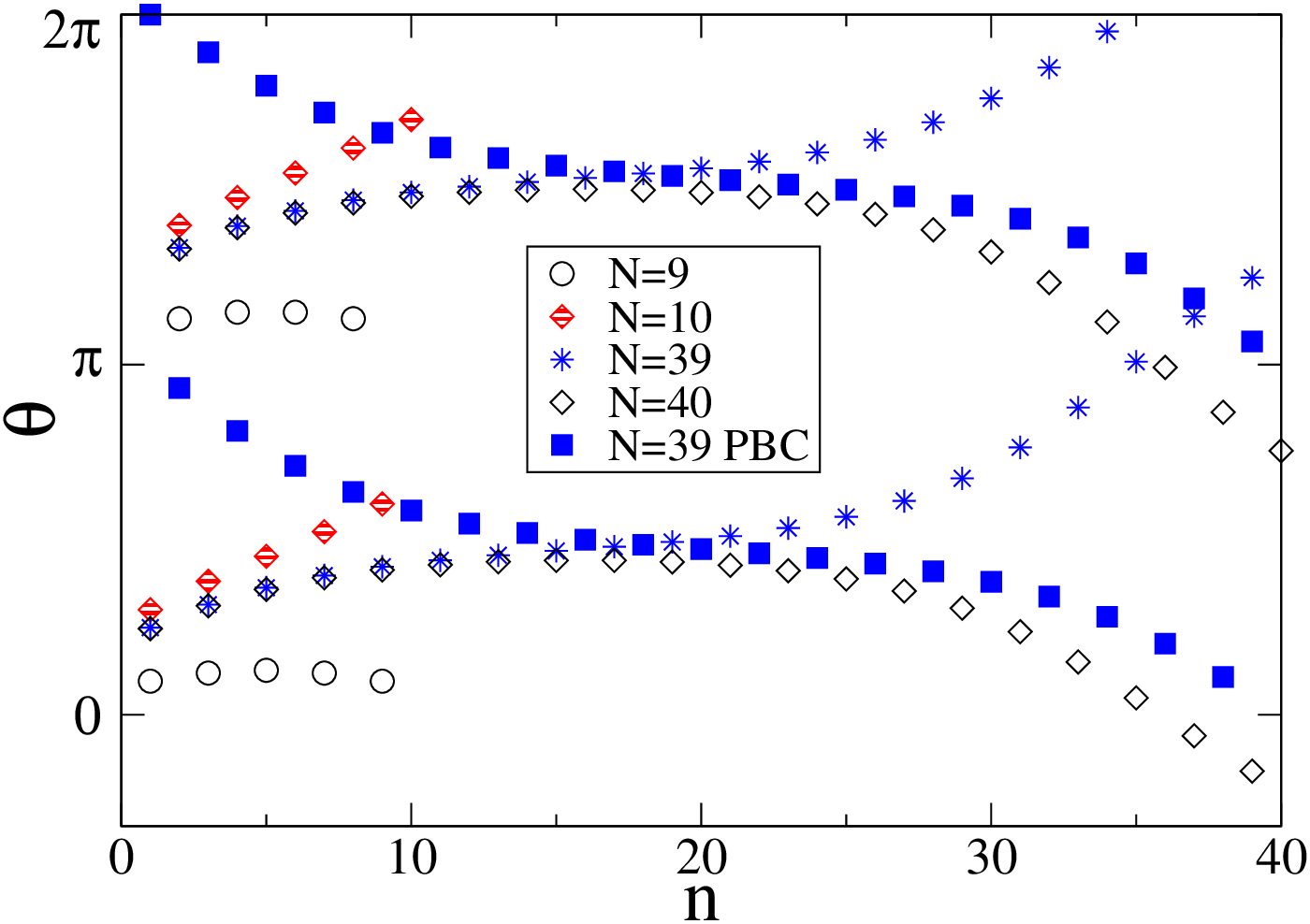}
\caption{ (color online) Equilibrium configurations $\theta_1,\dots,\theta_N$
calculated by the 2D map method for open chains with
$N=9,10,39,40$, and for a closed chain (PBC) 
with $N=39$.} \label{fig_config}
\end{figure}

Boundary conditions for open chains of $N$ atoms are taken into
account~\cite{PRL,Pandit} by introducing a fictitious
$(N+1)$-th atom and imposing $s_1=0=s_{N+1}$. The determination of
the ground state therefore corresponds to finding the value
$\theta_1$ such that, iterating the map $N$ times from the point
$P_1=(\theta_1,0)$, we get a point $P_{N+1}=(\theta_{N+1},0)$,
with both $P_1$ and $P_{N+1}$ located on the horizontal axis,
$s=0$. The $N$ values $\theta_1,\dots,\theta_N$ then give the
sought-after equilibrium configuration. In Fig.~\ref{fig_map} we
also plot the first $N$ steps of the map evolution giving the
ground states for $N=9$ (red solid squares) and $N=10$ (blue solid
circles). Different behaviors for even and odd $N$ can be inferred
from the different location of their trajectories in the phase portrait.
The configurations are explicitly shown in Fig.~\ref{fig_config}.

\begin{figure}
\includegraphics[width=8cm,clip=yes]{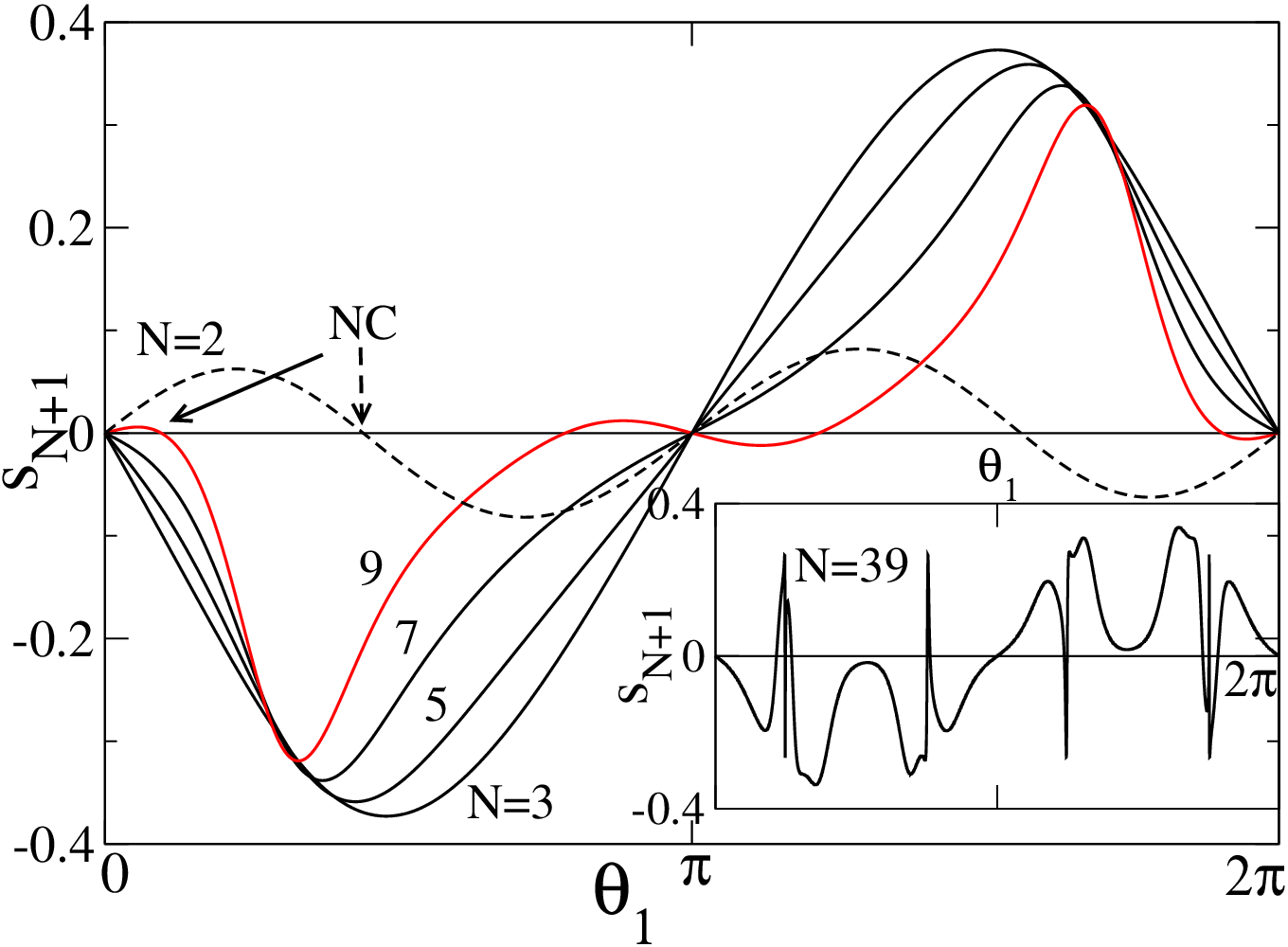}
\caption{ (color online) We plot $s_{N+1}(\theta_1)$, assuming
$s_1=0$, for different, small values of $N$ (main figure) and for
$N=39$ (inset). The arrows point to the equilibrium values
$\theta_1$ for the first atom of the $N=9$ chain (red full line, full
arrow) and of the $N=2$ chain (dashed line, dashed arrow).}
\label{fig_transition}
\end{figure}

The existence of a minimum length to get a noncollinear
configuration for odd $N$ is clear from Fig.~\ref{fig_transition},
where we plot $s_{N+1}$ as a function of $\theta_1$, assuming
$s_1=0$. For odd $N<9$, the only zeros are the AF fixed points,
corresponding to a collinear ferrimagnetic (FI) configuration, but
for $N=9$, $d s_{N+1}/d\theta_1$ changes sign at $\theta_1=0$, and
an additional solution appears: the noncollinear (NC)
configuration. For even $N$, non trivial solutions exist already
for $N=2$ (dashed line).
The Inset of Fig.~\ref{fig_transition} shows that for large $N$
the function $s_{N+1}(\theta_1)$ is strongly oscillating with
several zeros $\theta_1^{(k)}$. 
In order to determine the ground state,
the energies of all the NC
configurations with $\theta_1=\theta_1^{(k)}$ must be compared. 

We are now going to show that, by linearizing the map nearby the
fixed points, it is indeed possible to determine the exact {\it
analytical} condition for the rising of the NC state in the case
of an open chain with odd $N$. If we start from a point 
$P_1=(s_1,\theta_1)$ close to the fixed point $(0,0)$, iterated
points on the map are oscillating between the AF fixed points: 
more precisely, even $P_{2k}$ are close to $(\pi,0)$ and odd 
$P_{2k+1}$ are close to $(0,0)$. So, if we write
\bea
\theta_{2k} &=& \pi + \delta_{2k} \\
\theta_{2k+1} &=& \delta_{2k+1}
\eea
the quantities $\delta_n$ are small for any $n$, as $s_n$ are.
Now, we can linearize the map in the two cases $P_{2k-1} \to P_{2k}$
and $P_{2k} \to P_{2k+1}$. If we write
$$
\left( \begin{array}{c} \delta_{2k} \\ s_{2k} \end{array} \right)
= A_1 
\left( \begin{array}{c} \delta_{2k-1} \\ s_{2k-1} \end{array} \right)
\hbox{~~and~~}
\left( \begin{array}{c} \delta_{2k+1} \\ s_{2k+1} \end{array} \right)
= A_2 
\left( \begin{array}{c} \delta_{2k} \\ s_{2k} \end{array} \right)
$$
we get the matrices
$$
A_1 = \left( \begin{array}{cc} 
1+h & -1 \\
-h & 1
\end{array} \right)
\hbox{~~and~~}
A_2 = \left( \begin{array}{cc} 
1-h & -1 \\
h & 1
\end{array} \right) .
$$

So, if $N=2N_0+1$ is an odd integer, we have
$$
\left( \begin{array}{c} \delta_{N+1} \\ s_{N+1} \end{array} \right)
= A_1 (A_2 A_1)^{N_0} 
\left( \begin{array}{c} \delta_{1} \\ s_{1} \end{array} \right)
\equiv T
\left( \begin{array}{c} \delta_{1} \\ s_{1} \end{array} \right) .
$$
Since $s_{N+1}=T_{21}\delta_1 + T_{22} s_1$, if we start with $s_1=0$, the
condition $ds_{N+1}/d\theta_1=ds_{N+1}/d\delta_1=0$ 
reads $T_{21}=0$. Let us now implement this condition,
firstly determining eigenvalues $\lambda_i$ and eigenvectors $v_i$ of the matrix
$$
A = A_2 A_1 =
\left( \begin{array}{cc}
1+h-h^2 & h-2 \\
h^2 & 1-h
\end{array} \right) .
$$

It is easily found that
\bea
\lambda_{1,2} &=& \fra{1}{2}(2-h^2 \pm ih\sqrt{4-h^2}) \\
v_j &=& \left( \begin{array}{c}
{\lambda_j + h -1 \over h^2} \\ 1 \end{array} \right) \equiv
\left( \begin{array}{c} v_{1j} \\ 1 \end{array} \right) 
~~~ j=1,2 . 
\eea

If $U$ is the $(2\times2)$ matrix with $v_{1,2}$ as column vectors and
$A_D$ is the diagonal matrix with elements $\lambda_{1,2}$, it is
straightforward to write $T = A_1 U A_D^{N_0} U^{-1}$. 
Finally, the condition $T_{21} =0$ gives
\be
h(v_{11}\lambda_1^{N_0} - v_{12}\lambda_2^{N_0}) =
\lambda_1^{N_0} - \lambda_2^{N_0} ,
\ee
which simplifies to 
\be
\left( {\lambda_1 \over \lambda_2} \right)^{N_0} = {\lambda_2 -1
\over \lambda_1 -1} .
\ee

\begin{figure}
\includegraphics[width=8cm,clip=yes]{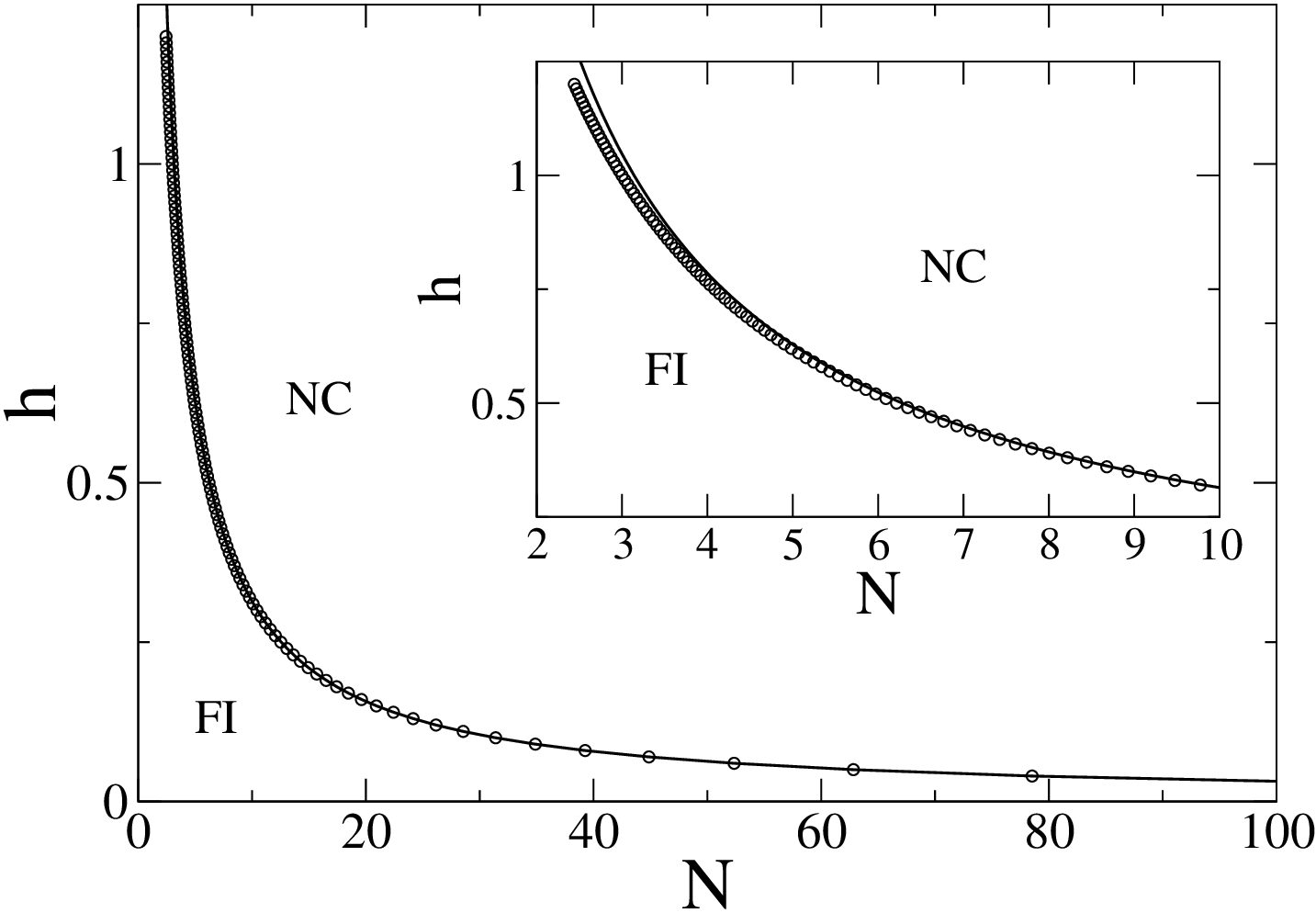}
\caption{ Main: analytical results of the 
phase diagram for odd-$N$ open chains, 
Eqs.~(\ref{eq_Nc1}-\ref{eq_Nc2}). FI and NC
denote FerrImagnetic and NonCollinear states. The full line is the
curve $h_c = \pi/N$ which is a very good approximation, even for
small $N$ (Inset). } \label{fig_hNdiagram}
\end{figure}

Therefore, the transition length is equal to $N_c=2N_0+1$ where $N_0$
is the solution of the above equation. We get
\be
N_c={\pi\over \varphi},
\label{eq_Nc1}
\ee
with 
\be
\cos\varphi= {2-h^2\over 2} ~~~~~ \sin\varphi={h\sqrt{4-h^2}\over 2} .
\label{eq_Nc2}
\ee 
The curve is plotted as circles in Fig.~\ref{fig_hNdiagram} along
with the asymptotic form $N_c=\pi/h$ (full line) which appears to
be a very good approximation even for small $N$ (see the Inset).

We now turn to closed chains, which imply periodic boundary
conditions (PBC).
If spins represent magnetic layers, these
boundary conditions are not physical, but for nanowires
deposited on a substrate they are physical and correspond to nanorings.
 In terms of the 2D mapping, PBC imply 
$P_{N+1}\equiv P_1$, i.e. $\theta_{N+1}=\theta_1$ and
$s_{N+1}=s_1$. Therefore, trajectories are fixed points
of order $N$. It is easy to realize that the ground state
for even $N$ is the bulk spin-flop state: 
$\theta_{2k}=\bar\theta$ and $\theta_{2k+1}=-\bar\theta$,
with $\cos\bar\theta=h/4$.
In fact, if $\theta_1,\dots,\theta_N$ were a different
configuration with a lower energy, we might replicate it
indefinitely for an infinite chain and get a configuration
with an energy lower than the bulk spin-flop phase (which
is the ground state).

\begin{figure}
\includegraphics[width=6cm,clip=yes]{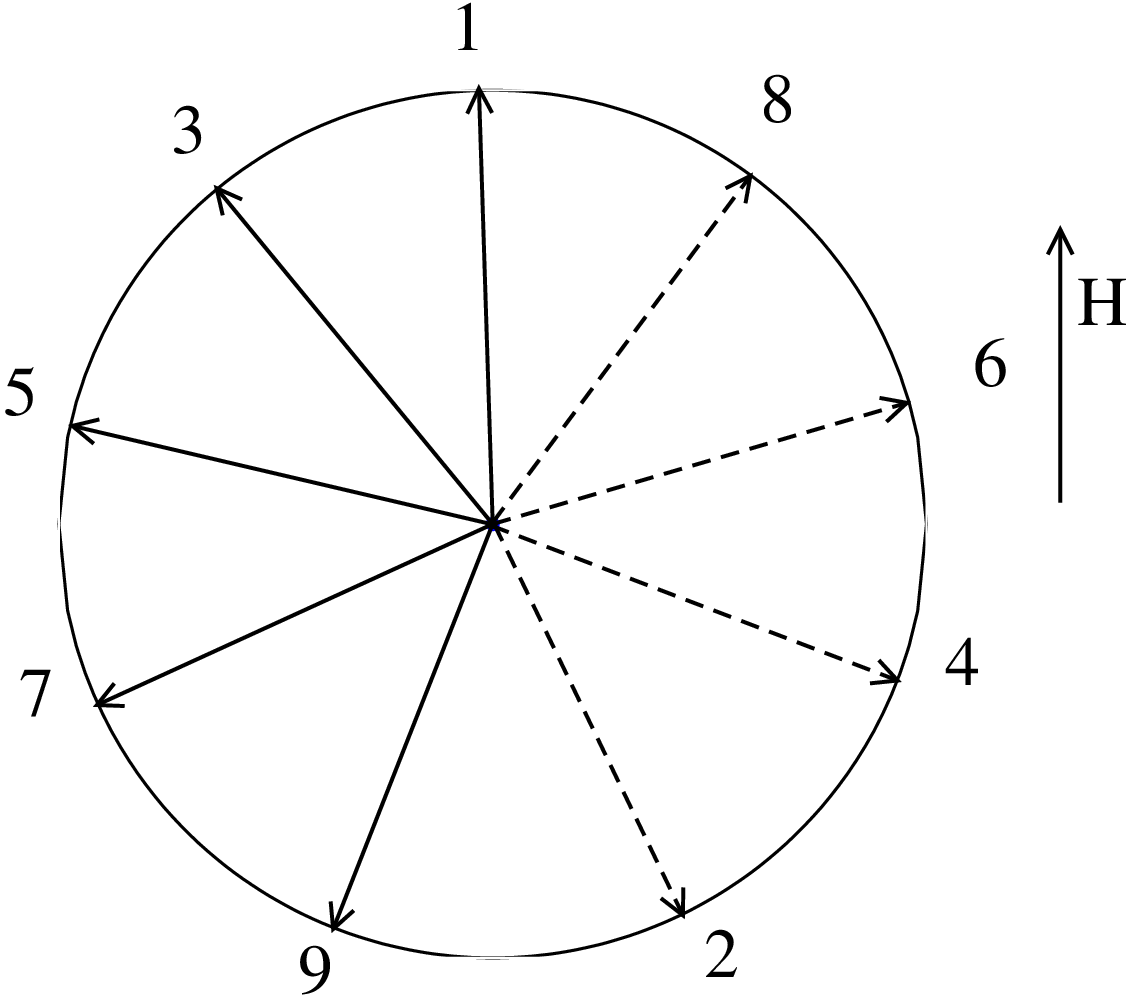}
\caption{ Graphical representation of the ground state for a
closed chain with $N=9$.
Full and dashed arrows represent spin orientations on odd
and even sites, respectively. $\theta_1=0$ and for every
angle $\theta_{2k}$ there is an angle $\theta_{2j+1}=-\theta_{2k}$.} 
\label{fig_9periodic}
\end{figure}

The above argument does not apply to odd $N$, because
the SF phase, as well as the AF phase, which are
fixed points of order two, do not satisfy PBC for odd $N$.
In this case, we have a nonuniform state. In order to
find it, we should look for the points $P$ that are 
iterated on themselves after $N$ applications of the map.
It appears that the configuration is symmetric with respect 
to the field direction, i.e. for any spin with an angle
$\theta^*$ there is a spin forming an angle $-\theta^*$,
see Fig.~\ref{fig_9periodic} for $N=9$. Therefore, for
odd $N$, there is one spin with $\theta=0$. PBC allow to label
this spin as ``number 1", so that searching the solution
is now as easy as for open chains: 
we apply the map $N$ times to points $(0,s_1)$ and look for
the values $s_1$ such that $\theta_{N+1}=0$ and
$s_{N+1}=s_1$.
Using this method, we have found the ground state for $N=39$
(Fig.~\ref{fig_config}, full squares) and for $N=9$
(Fig.~\ref{fig_9periodic}).

In conclusion, we have studied model (\ref{competing})
which describes a chain of classical planar spins with nearest
neighboring AF coupling and interacting with a (real or effective)
external field $h$. Since the anisotropy is zero, the infinite system is
in the bulk SF phase for any nonvanishing $h$. Extrema of
the energy correspond to trajectories of the 2D mapping
(\ref{mappa}) with appropriate boundary conditions:
$s_1=s_{N+1}=0$ for open chains and
$\theta_{N+1}=\theta_1,s_{N+1}=s_1$ for closed chains.

This method allows a fast and exact determination of the ground states
for any $N$ (Fig.~\ref{fig_config}).
It also allows to find analytically the transition length
$N_c$ for open odd chains from the FI to the NC state (Fig.~\ref{fig_hNdiagram}).
This transition corresponds to a change in sign of the
derivative $ds_{N+1}/d\theta_1$ at $\theta_1=0$ (Fig.~\ref{fig_transition}).
Parity effects are present for open and closed chains.

With increasing the field $h$, the 2D map starts developing
a chaotic behavior~\cite{Bogdanov,chaos}. 
Studying this regime with reference to nanowires would be
an interesting subject for future work.


\begin{thebibliography}{99}

\bibitem{Neel}
L. N\'eel, Ann. Phys. (Paris) {\bf 5}, 232 (1936).

\bibitem{Mills}
D. L. Mills, Phys. Rev. Lett. {\bf 20}, 18 (1968).

\bibitem{WangMillsPRL}
R. W. Wang, D. L. Mills, E. E. Fullerton, J. E. Mattson, and S. D.
Bader, Phys. Rev. Lett. {\bf 72}, 920 (1994).

\bibitem{WangMillsPRB}
R. W. Wang and D. L. Mills, Phys. Rev. B {\bf 50}, 3931 (1994).

\bibitem{PRL}
L. Trallori, P. Politi, A. Rettori, M. G. Pini, and J. Villain, 
Phys. Rev. Lett. {\bf 72}, 1925 (1994).

\bibitem{JAP}
L. Trallori, P. Politi, A. Rettori, M. G. Pini, and J. Villain, 
J. Appl. Phys. {\bf 76}, 6555 (1994).

\bibitem{IJMPB}
L. Trallori, M. G. Pini, A. Rettori, M. Macci\`o, and P. Politi, 
Int. J. Mod. Phys. B {\bf 10}, 1935
(1996).

\bibitem{Bogdanov}
U. K. R\"ossler and A. N. Bogdanov,
Phys. Rev. B {\bf 69}, 094405 (2004).

\bibitem{Science}
C. F. Hirjibehedin, C. P. Lutz, and A. J. Heinrich, Science {\bf
312}, 1021 (2006).

\bibitem{LounisPRB}
S. Lounis, Ph. Mavropoulos, P. H. Dederichs, and S. Bl\"ugel,
Phys. Rev. B {\bf 72}, 224437 (2005).

\bibitem {LounisPRL}
S. Lounis, P. H. Dederichs, and S. Bl\"ugel,
Phys. Rev. Lett. \textbf{101}, 107204 (2008).

\bibitem{notaLounis}
Notice that frustration occurs also in the case of {\it
antiferromagnetic} exchange coupling between an atom of the AF
wire and a magnetic moment of the ferromagnetic substrate, as is
the case of AF Cr wires deposited on Ni(001) \cite{LounisPRB}.

\bibitem{Aubry}
S. Aubry, in {\it Solitons and Condensed Matter Physics},
edited by A. R. Bishop and T. Schneider (Springer, 1979).

\bibitem{Bak}
P. Bak, Phys. Rev. Lett. {\bf 46}, 791 (1981).

\bibitem{Belorov}
P. I. Belorov {\it et al.}, Zh. Eksp. Teor. Fiz. {\bf 87},
310 (1984) [Sov. Phys. JETP {\bf 60}, 180 (1984)].

\bibitem{Pandit}
R. Pandit and M. Wortis, Phys. Rev. B {\bf 25}, 3226 (1982).

\bibitem{note_sin}
The function $\sin^{-1}$ is a two-values function.
The correct value~\cite{Belorov} is within $\pi/2$ and $3\pi/2$.

\bibitem{Ott}
E. Ott, {\it Chaos in Dynamical Systems}
(Cambridge University Press, 2002).

\bibitem{chaos}
L. Trallori, P. Politi, A. Rettori, M.G. Pini, and J. Villain,
J. Phys.: Cond. Matt. {\bf 7}, L451 (1995).

\end{thebibliography}
\end{document}